\documentclass[conference]{IEEEtran}
\IEEEoverridecommandlockouts
\usepackage{cite}
\usepackage[cmex10]{amsmath}
\usepackage{array}
\usepackage{amsfonts}
\usepackage{mathrsfs}
\usepackage{arydshln}
\usepackage{slashbox}
\usepackage{graphicx}
\usepackage{float}

\usepackage{subfigure}
\usepackage{amssymb}
\usepackage{amsmath}
\usepackage{enumitem}
\usepackage[colorlinks,linkcolor=black,urlcolor=black,anchorcolor=black,citecolor=black,hyperfootnotes=true]{hyperref}
\usepackage{multirow}
\usepackage{multicol}
\usepackage{cite}
\usepackage{url}
\usepackage[subfigure]{graphfig}
\usepackage{xcolor}
\usepackage{setspace}
\newcommand{\mv}[1]{\mbox{\boldmath{$ #1 $}}}
\usepackage[linesnumbered,ruled]{algorithm2e}

\newtheorem{proposition}{\underline{Proposition}}

\newcommand{\qed}{\nobreak \ifvmode \relax \else
\ifdim\lastskip<1.5em \hskip-\lastskip
\hskip1.5em plus0em minus0.5em \fi \nobreak
\vrule height0.75em width0.5em depth0.25em\fi}
{\setlength\abovedisplayskip{0.4pt}
\setlength\belowdisplayskip{0.4pt}
\newcommand{\cc}{\mathrm{C}}

\begin{document}
\title{Trajectory Optimization of Cellular-Connected UAV for Information Collection and Transmission}	
\author{\IEEEauthorblockN{Xianzhen Guo$^{\dagger}$, Shuowen Zhang$^{\star}$$^{\dagger}$, and Liang Liu$^{\star}$$^{\dagger}$\\
		$^{\star}$Department of Electronic and Information Engineering, The Hong Kong Polytechnic University\\
		$^{\dagger}$Shenzhen Research Institute, The Hong Kong Polytechnic University\\
		E-mails: xianzhen.guo@connect.polyu.hk, \{shuowen.zhang, liang-eie.liu\}@polyu.edu.hk}
\vspace{-6mm}
\thanks{This work was supported by the National Natural Science Foundation of China under Grant 62101474. This work was also supported by Shenzhen Virtual University Park Management Center under Grant R2021A006.}}
\vspace{-1mm}
\maketitle 

\begin{abstract}
In this paper, we consider a cellular-connected unmanned aerial vehicle (UAV) with an information collection and transmission mission for multiple ground targets. Specifically, the UAV is required to collect a fixed amount of information of each target by hovering at a pre-determined location (via e.g., photography/videography/sensing), and transmit all the collected information to the cellular network during its flight. We aim to jointly optimize the UAV's trajectory and the information collection order of the ground targets to minimize the mission completion time. The formulated problem is NP-hard due to the need of visiting the information collection locations for all targets; moreover, the UAV's trajectories over different time durations are coupled in non-convex constraints for ensuring  information transmission completion. To handle this difficult problem, we first propose a structured communication protocol between the UAV and the cellular network, which decouples the UAV's trajectory designs in different time durations. Then, under the proposed protocol, we establish an equivalent graph-based model for the considered problem, and devise a low-complexity algorithm for finding an approximate solution by exploiting the problem structure and leveraging graph theory. Numerical results show that our proposed design achieves efficient information collection and transmission, and \hbox{outperforms various benchmark schemes.} 
\end{abstract}
\vspace{-5mm}
\section{Introduction}
\vspace{-2mm}
Unmanned aerial vehicles (UAVs) have found numerous interesting applications, due to its highly controllable mobility and swift deployment \cite{Access}. To support the various missions of UAVs, it is of paramount importance to realize high-quality UAV-to-ground communications. A promising technology to achieve this goal is \emph{cellular-enabled UAV communication} \cite{Cellular_Enabled}, or \emph{cellular-connected UAV}, where UAVs are integrated to the cellular network as new aerial users and served by the ground base stations (GBSs). Unlike existing Wi-Fi based technologies, cellular-enabled UAV communication enables beyond visual line-of-sight (LoS) communication range, low latency, and ubiquitous accessibility by exploiting the high-speed backhaul links among the GBSs in the cellular network.

To maximally harness the benefits of cellular-enabled UAV communication, the UAV's trajectory needs to be judiciously designed to ensure satisfactory communication performance during its flight. Note that the information communicated between the UAVs and the GBSs can be categorized into two types: low-rate \emph{control information} for ensuring the safety of UAVs, and high-rate \emph{payload information} when the UAV needs to acquire information and deliver it to the cellular network (e.g., in surveillance and inspection applications where the UAV needs to take images, record videos, or perform sensing of critical infrastructures) \cite{Cellular_Enabled}. Existing works on trajectory optimization of cellular-connected UAVs mostly considered control information transmission, where the instantaneous quality-of-service (QoS) needs to meet certain requirement during the flight for ensuring reliable control of the UAVs, see, e.g., \cite{Cellular_Enabled,interference,outage,outage1,Gesbert,federated,radio}. On the other hand, for payload information transmission, the communication performance metric is generally the sum achievable rate or total transmitted data volume, which makes the optimal trajectory drastically different from that for control information transmission, e.g., the UAV may need to fly in high-rate regions for longer time instead of only in GBS coverage regions with minimum required QoS. The trajectory optimization considering payload information transmission is still in its infancy with few studies, e.g., \cite{Zhan} which aimed to maximize the sum rate over the UAV's flight.

\begin{figure}[t]
\centering
\includegraphics[width=5cm]{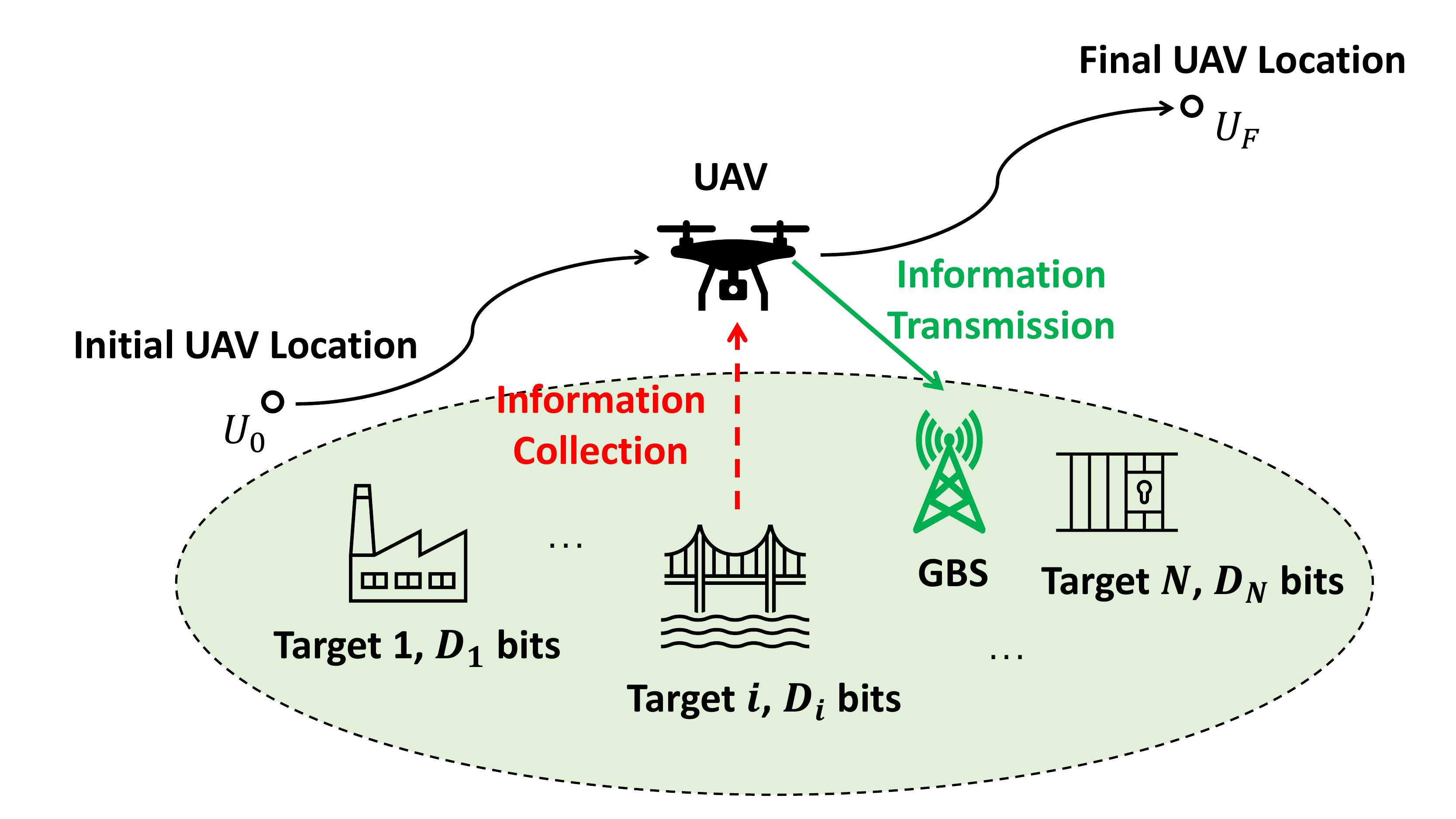}
\vspace{-3mm}
\caption{A cellular-connected UAV for information collection and transmission.}\label{system_model}
\vspace{-3mm}
\end{figure}

In this paper, we consider a novel and practical \emph{information collection and transmission mission} of a cellular-connected UAV with multiple ground targets, as illustrated in Fig. \ref{system_model}, which has not been investigated before to the best of our knowledge. Specifically, the UAV needs to collect a given amount of information of each target while hovering at a pre-determined information collection point (e.g., via photography/videography/sensing) for a given amount of time, and transmit the collected information to the GBS during its flight. Our objective is to minimize the mission completion time of the UAV, by jointly optimizing the UAV's trajectory and the information collection order. This problem is NP-hard since it can be reduced to a traveling salesman problem (TSP) due to the need of visiting all the information collection points. Moreover, the problem also involves non-convex constraints that couple the UAV's trajectories in different time durations. To tackle this problem, we devise a structured communication protocol to decouple these constraints, based on which we develop an equivalent graph-based model of the problem. By exploiting the problem structure, we propose a low-complexity suboptimal solution to the problem. Numerical results demonstrate the superiority of our proposed solution compared to various benchmark schemes based on the traditional TSP method or successive convex approximation (SCA).

\vspace{-1mm}
\section{System Model}
\vspace{-1mm}
Consider an information collection and transmission mission of a cellular-connected UAV, as illustrated in Fig. \ref{system_model}. The UAV is required to collect information of $N\!\geq\! 1$ ground targets (e.g., critical infrastructure) by hovering at given locations, and transmit the collected information to a GBS in the cellular network during its flight. The information collection and transmission processes are modeled as follows.
\vspace{-1mm}
\subsection{Information Collection}
\vspace{-1mm}
Denote $\mathcal{N}=\{1,...,N\}$ as the set of ground targets. To collect the information of each $i$-th target, $i\in \mathcal{N}$, the UAV is required to hover at a given \emph{information collection point} denoted by $A_i$ for $T^{\mathrm{C}}_i$ seconds (s). The volume of information to be collected for target $i$ is fixed as $D_i$ bits. Note that $A_i$'s, $T^{\mathrm{C}}_i$'s, and $D_i$'s are pre-determined based on the target properties and information collection methods. For example, $A_i$'s can be set as the optimal shooting/sensing positions that achieve the best image/video/sensing quality; $T^{\mathrm{C}}_i$'s can be set as the required time to complete shooting/sensing of the targets; and $D_i$'s can be determined based on the required image/video size and resolution or sensing accuracy.

For ease of exposition, we consider homogeneous targets with a common height of the information collection points $A_i$'s denoted by $H_U$ meters (m), and further assume that the UAV flies at a constant altitude $H_U$.\footnote{The results in this paper can be readily extended to the case with heterogeneous target information collection heights and adjustable UAV altitude.} Let $\mv{u}(t)\!\in\! \mathbb{R}^{2\times 1}$ denote the UAV's horizontal location at time instant $t$, thus the UAV's trajectory projected on the horizontal plane can be expressed as $\{\mv{u}(t),t\!\in\! [0,T]\}$, where $T$ denotes the \emph{mission completion time} of the UAV. We further consider a set of given initial and final points of the UAV denoted by $U_0$ and $U_F$ with horizontal locations $\mv{u}_0\!\in\! \mathbb{R}^{2\times 1}$ and $\mv{u}_F\!\in\! \mathbb{R}^{2\times 1}$, respectively, which yields $\mv{u}(0)\!=\!\mv{u}_0$ and $\mv{u}(T)\!=\!\mv{u}_F$. The maximum speed of the UAV is denoted as $V_{\max}$, i.e., $\|\dot{\mv{u}}(t)\|\!\!\leq\!\! V_{\max},\forall t\!\in\! [0,T]$. Moreover, denote $\mv{a}_i\!\in\! \mathbb{R}^{2\times 1}$ as the horizontal location of the information collection point for the $i$-th target, $\mv{g}\!\in\! \mathbb{R}^{2\times 1}$ as the horizontal location of the GBS, and $H_G$ as the height of the GBS.

\begin{figure}[t]
\vspace{1mm}
\centering
\includegraphics[width=8.5cm]{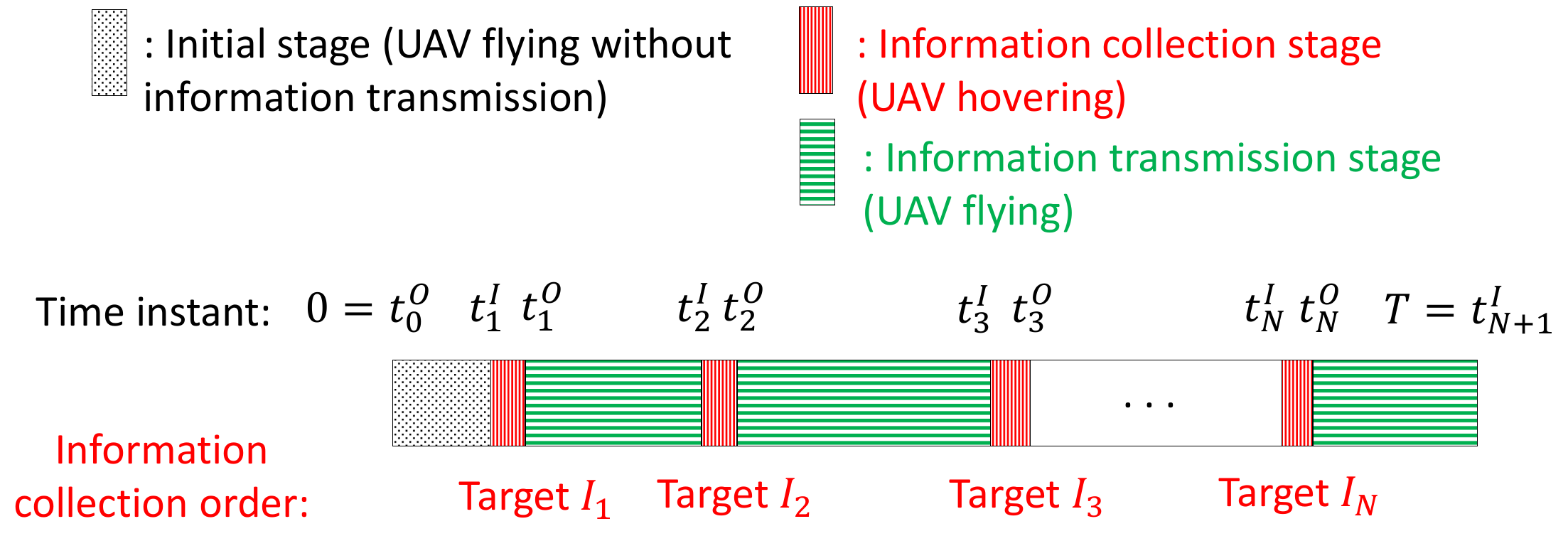}
\vspace{-3mm}
\caption{Illustration of different stages in the UAV's mission.}
\label{protocol}
\vspace{-3mm}
\end{figure}

Note that to collect the information of all the $N$ targets, the UAV needs to visit the $N$ information collection points $A_i$'s one-by-one. Let $\mv{I}=[I_1,...,I_N]^T$ denote the \emph{information collection order}, where $I_i\!\in\! \mathcal{N}$ denotes the index of target whose information collection point $A_{I_i}$ is visited in the $i$-th order, with $\{I_i\}_{i=1}^N=\mathcal{N}$. We further denote $t_i^I$ and $t_i^O$ as the critical time instants when the UAV starts or stops its hovering at $A_{I_i},i\!\in\! \mathcal{N}$ for information collection, respectively, where $t_i^O\!-\!t_i^I\!=\!T_{I_i}^{\mathrm{C}}$. For ease of notation, we further define $t_0^O=0$ and $t_{N+1}^I=T$. Thus, the UAV's trajectory should satisfy 
\begin{align}\label{collection}
\mv{u}(t)=\mv{a}_{I_i},\quad t\in[t_i^I,t_i^O],\ \forall i\in \mathcal{N}.
\end{align}
In Fig. \ref{protocol}, we illustrate the information collection order and critical time instants. It can be observed that there are $N$ \emph{information collection stages} in the UAV's mission, where each $i$-th stage serves target $I_i$ in time period $\mathcal{T}_i^{\cc}=[t_i^I,t_i^O]$.
\vspace{-1mm}	
\subsection{Information Transmission}
\vspace{-1mm}
Between two consecutive information collection stages indexed by $i$ and $i+1$, the UAV flies from information collection point $A_{I_i}$ to $A_{I_{i+1}}$ while transmitting the collected information to the GBS during its flight until all the information has been delivered, as illustrated in Fig. \ref{protocol}.\footnote{We assume information collection and information transmission are scheduled in orthogonal time periods, i.e., no information is transmitted when the UAV is hovering at information collection points.} It can be observed from Fig. \ref{protocol} that there are $N$ \emph{information transmission stages} in the UAV's mission, where each $i$-th stage takes time period  $\mathcal{T}_i^{\mathrm{T}}=[t_i^O,t_{i+1}^I]$. Note that if the information of a target $I_i$ is not fully transmitted in the $i$-th information transmission stage due to limited rate/time, it is stored in the UAV and can be transmitted in the following $(i+1)$-th to $N$-th stages.

We assume that both the UAV and the GBS are equipped with an omni-directional antenna with unit gain. The distance between the UAV and the GBS at time instant $t$ is expressed as $d(\boldsymbol{u}(t)) = \sqrt{\|\boldsymbol{u}(t)-\boldsymbol{g}\|^2+H^2}$, where $H=H_U-H_G$. For the purpose of drawing essential insights, we assume that the UAV-GBS communication channel is dominated by the LoS link.\footnote{Note that for surveillance applications considered in this paper, the UAV usually flies in rural and remote areas without significant scatters, where the LoS model can characterize the UAV-GBS channel accurately.} Thus, the channel power at time instant $t$ can be expressed as follows under the free-space path loss model:
\vspace{0.5mm}\begin{equation}
h^2(\boldsymbol{u}(t)) = \beta_0/d^2(\boldsymbol{u}(t))=\beta_0/(\|\boldsymbol{u}(t)-\boldsymbol{g}\|^2+H^2),
\vspace{0.5mm}\end{equation}
where $\beta_0$ denotes the channel power gain at the reference distance $d_0 = 1$ m. Let $B$ denote the bandwidth allocated to UAV-GBS information transmission in Hertz (Hz), $P$ denote the transmission power at the UAV, and $\sigma^2$ denote the average noise power at the GBS receiver. The achievable rate from the UAV to the GBS at time instant $t$ can be expressed as
\vspace{0.5mm}\begin{equation}\label{rate}
\!\!	R(\boldsymbol{u}(t))\!=\! 
\begin{cases}
	B{\rm log}_2\left(1\!+\!\frac{P\beta_0/\sigma^2}{\|\boldsymbol{u}(t)-\boldsymbol{g}\|^2+H^2}\right),\!\!\!\!\! &t\!\in\! \mathcal{T}_i^{\mathrm{T}}, i\!\in\! \mathcal{N},\\[-1mm]
	0, &t\notin {\cup}_{i=1}^N\mathcal{T}_i^{\mathrm{T}},
\end{cases}
\vspace{0.5mm}\end{equation}
which is non-zero only in the information transmission stages.

Since the information of each $I_i$-th target is collected by the UAV at time instant $t_i^O$, it can only be transmitted after $t_i^O$ during the $i$-th to the $N$-th information transmission stages denoted by $\cup_{j=i}^N\mathcal{T}_j^{\mathrm{T}}=\cup_{j=i}^N[t_j^O,t_{j+1}^I]$. Therefore, it can be easily proved by induction that ensuring successful information transmission for all targets before mission completion is equivalent to guaranteeing that for any $i\in \mathcal{N}$, the total \emph{transmitted} information volume in the $i$-th to the $N$-th information transmission stages is no smaller than the total \emph{collected} information volume in the $i$-th to the $N$-th information collection stages, i.e.,
\vspace{0.5mm}\begin{equation}
\sum_{j=i}^{N}\int_{t_j^O}^{t_{j+1}^I}R(\mv{u}(t))dt
\geq \sum_{j=i}^{N}D_{I_j},\quad \forall i \in \mathcal{N}.\label{info_causality}
\end{equation}

\subsection{Summary of the UAV's Mission}
In Fig. \ref{protocol}, we summarize the UAV's overall mission, which is divided into an initial stage $[0,t_{1}^I]$, $N$ information collection stages, and $N$ information transmission stages. Specifically, in the initial stage, the UAV flies to the firstly visited information collection point $A_{I_1}$ without information transmission, since no information has been collected. Then, the UAV sequentially collects information of the $N$ targets in $N$ stages with order $\mv{I}$, and performs information transmission during its flight. The overall mission completion time $T$ can be thus expressed as the sum duration of these stages:
\begin{align}\label{time}
T=\sum_{i=1}^N T_i^{\cc}+\sum_{i=0}^N(t_{i+1}^I-t_i^O).
\end{align}

Note that to meet the information collection and transmission constraints in (\ref{collection}) and (\ref{info_causality}) with minimum mission completion time $T$, the UAV's trajectory $\{\mv{u}(t),t\in [0,T]\}$ needs to be judiciously designed together with the order $\mv{I}$, based on the locations of the ground targets and GBS, as well as the required information volume for each target. For example, if a target with a large information volume is served lastly, the UAV may need to spend a long time in the last information transmission stage, which prolongs the mission completion time. As another example, if a target far away from the initial point is served firstly, the UAV needs to spend a long time flying to that target without any information transmission in the initial stage. Therefore, how to jointly optimize the trajectory and information collection order is a non-trivial problem, which will be investigated in this paper.
\vspace{-1mm}
\section{Problem Formulation}
\vspace{-1mm}
In this paper, we aim to optimize the trajectory of the cellular-connected UAV to minimize the completion time of the information collection and transmission mission. By introducing the information collection order $\mv{I}$ and critical time instants $\{t_i^I, t_i^O\}_{i=1}^{N}$ as auxiliary optimization variables, the optimization problem is formulated as
\begin{align}
\!\!\!\!\!	\mbox{(P1)}\!\!\underset{\scriptstyle \{\mv{u}(t),t\in [0,T]\},
	\atop \scriptstyle	\boldsymbol{I},T,\{t_i^I, t_i^O\}_{i=1}^{N}}{\min}\!\! &T \\[-0.5mm]
\mbox{s.t.}\quad 
&\sum_{j=i}^{N}\int_{t_j^O}^{t_{j+1}^I}R(\mv{u}(t))dt
\geq \sum_{j=i}^{N}D_{I_j}, \forall i \in \mathcal{N} \label{P1c_info}\\[-0.5mm]
& \boldsymbol{u}(t) = \boldsymbol{a}_{I_i}, \qquad\quad\  \forall t\in[t_i^I,t_i^O],\forall i\in \mathcal{N} \label{P1c_point}\\[-0.5mm]
& \|\dot{\boldsymbol{u}}(t)\| \leq V_{{\rm max}},\qquad\qquad\qquad  \forall t\in [0,T] \label{P1c_speed}\\[-0.5mm]
& \{I_i\}_{i=1}^N=\mathcal{N}\\[-0.5mm]
& \boldsymbol{u}(0) = \boldsymbol{u}_{0}\\[-0.5mm]
&\boldsymbol{u}(T) = \boldsymbol{u}_{F} \label{P1c_final}.
\end{align}

Problem (P1) is a non-convex optimization problem since the constraints in (\ref{P1c_info}) can be shown to be non-convex, which also involve integrals of the rate function that are difficult to be expressed explicitly. Particularly, note that the UAV's sub-trajectories $\{\mv{u}(t),t\in [t_i^O,t_{i+1}^I]\}$'s in different information collection stages are coupled in (\ref{P1c_info}). Moreover, the UAV's continuous trajectory $\{\mv{u}(t),t\in [0,T]\}$ involves an infinite number of optimization variables, which makes (P1) more challenging to solve.  On the other hand, (P1) is a combinatorial optimization problem due to the discrete optimization variables in the information collection order $\mv{I}$. Note that for the special case with $D_{I_i}=0,\forall I_i\in \mathcal{N}$, (P1) is reduced to the classic TSP, which is NP-hard \cite{TSP1}. Therefore, (P1) is also an NP-hard problem, for which the \hbox{optimal solution is difficult to obtain.}

To overcome these challenges, we propose a low-complexity algorithm to find a suboptimal solution of (P1) as follows.
\vspace{-1mm}
\section{Proposed Solution to (P1)}
\vspace{-1mm}
In this section, we propose a graph theory based approach to find a high-quality suboptimal solution to (P1) with low complexity. Specifically, to resolve the difficulty resulted from the coupling among the sub-trajectories for different information transmission stages in (\ref{P1c_info}), we devise a transmission protocol with a simple structure to decouple these constraints and reformulate the problem. Next, we establish an equivalent graph-based model of the reformulated problem, and propose an efficient algorithm for finding a suboptimal solution.
\subsection{Reformulation of (P1)}
First, we reformulate (P1) to decouple the constraints on the sub-trajectories in different information transmission stages. Specifically, we propose a simple information transmission structure to fulfill the constraints in (\ref{P1c_info}), where the $D_{I_i}$-bit information of each $I_i$-th target needs to be \emph{fully transmitted} to the GBS during the $i$-th information transmission stage, before the UAV collects information of the next ($I_{i+1}$-th) target. Therefore, the constraints in (\ref{P1c_info}) can be automatically satisfied as long as the transmitted information volume (integrated rate) in each $i$-th information transmission stage is no smaller than $D_{I_i}$. Note that the UAV's trajectory can be equivalently characterized by the sub-trajectories in the initial stage and information transmission stages, $\{\mv{u}(t),t\in[t_i^O,t_{i+1}^I]\}_{i=0}^N$; moreover, it follows from (\ref{time}) that minimizing the mission completion time $T=t_{N+1}^I$ is equivalent to minimizing the total time in the initial stage and information transmission stages, $\sum_{i=0}^N(t_{i+1}^I-t_i^O)$. Therefore, (P1) can be reformulated as follows under the proposed protocol:
\begin{align}
\!\!\!\!\mbox{(P2)}\underset{\scriptstyle \{\mv{u}(t),t\in[t_i^O,t_{i+1}^I]\}_{i=0}^N
	\atop \scriptstyle t_{N+1}^I,\{t_i^I, t_i^O\}_{i=1}^{N},\boldsymbol{I}	}{\min}  &\sum_{i=0}^N(t_{i+1}^I-t_i^O)\\
\mbox{s.t.}\quad
& \int_{t_i^O}^{t_{i+1}^I}R(\mv{u}(t))dt  \!\geq \!D_{I_i}, \forall i \!\in\! \mathcal{N} \label{P2c_info}\\
& (\ref{P1c_point})-(\ref{P1c_final}).
\end{align}

Note that the sub-trajectories in different information transmission stages are no longer coupled in the new information transmission constraints (\ref{P2c_info}), thus making (P2) much more tractable. Moreover, any feasible solution to (P2) is a feasible solution to (P1). In the following, we focus on (P2). 

It is worth noting that (P2) is still an NP-hard problem since it can be reduced to a TSP similarly as (P1). Moreover, the discrete optimization variables in $\mv{I}$ makes (P2) difficult to be handled by standard optimization methods. Nevertheless, thanks to the decoupled constraints, we are able to establish an equivalent graph-based model for (P2), based on which efficient algorithms in graph theory can be leveraged to find a high-quality approximate solution, as elaborated below.
\vspace{-1mm}
\subsection{Equivalent Graph-Based Model for (P2)}
\vspace{-1mm}
In this subsection, we present an equivalent graph-based model for (P2). Note that for any given information collection order $\mv{I}$, (P2) aims to minimize the total time duration of the initial stage and $N$ information transmission stages by optimizing the sub-trajectories in these stages, while ensuring the transmitted information volume in each information transmission stage $i$ is no smaller than the previously collected information volume for the $I_i$-th target, $D_{I_i}$. Therefore, the optimal solution to (P2) for given $\mv{I}$ can be found by optimizing the sub-trajectories of the UAV in the initial stage and $N$ information transmission stages independently \emph{in parallel}. 

For ease of notation, define $I_0=0$, $I_{N+1}=N+1$, $\mv{a}_{I_0}=\mv{u}_0$, $\mv{a}_{I_{N+1}}=\mv{u}_F$, and $D_{I_0}=0$. Let $T(I_i,I_{i+1})=t_{i+1}^I-t_i^O$ denote the time duration for the initial stage (with $i=0$) or the $i$-th information transmission stage (with $i\in \mathcal{N}$), respectively, and $\{\bar{\mv{u}}(t),t\!\in\! [0,T(I_i,I_{i+1})]\}$ as the sub-trajectory therein. The problem to optimize each sub-trajectory is formulated as
\begin{align}
\!\!\!\!\!	\mbox{(P2-Sub)}\!\!\!\!\!\underset{\scriptstyle T(I_i,I_{i+1})\atop \scriptstyle \{\bar{\mv{u}}(t),t\in[0,T(I_i,I_{i+1})]\}}{\min}\!\!\!\!  & T(I_i,I_{i+1})\\[-0.5mm]
\mbox{s.t.}\quad
& \int_{0}^{T(I_i,I_{i+1})}R(\bar{\mv{u}}(t))dt  \geq D_{I_i}\label{P2subc_info} \\[-0.5mm]
& \bar{\mv{u}}(0) = \mv{a}_{I_i}\\[-0.5mm]
& \bar{\mv{u}}(T(I_i,I_{i+1})) = \mv{a}_{I_{i+1}}\\[-0.5mm]
& \|\dot{\bar{\mv{u}}}(t)\| \leq V_{{\rm max}},\quad \forall t.
\end{align}

Note that any feasible solutions of $T(I_i,I_{i+1})$'s and $\{\bar{\mv{u}}(t),t\!\in\![0,T(I_i,I_{i+1})]\}$'s to (P2-Sub) correspond to an equivalent feasible solution of (P2) below, and vice versa:
\begin{align}
t_{i+1}^{I}=&\sum_{j=1}^{i}T_{I_j}^{\cc}+\sum_{j=1}^{i+1}T(I_{j-1},I_{j}),\quad t_i^{O}=t_i^I+T_{I_i}^{\cc}\label{P2sol_1}\\[-0.5mm]
\mv{u}(t)=&
\begin{cases}
	\bar{\mv{u}}(t+t_{i-1}^{O}),t\in [t_{i-1}^{O},t_i^{I}]\\[-0.5mm]
	\mv{a}_{I_i},\qquad\quad\ t\in [t_{i}^{i},t_i^{O}]
\end{cases},\quad \forall i\in \mathcal{N}.\label{P2sol_2}
\end{align}
Thus, the optimal information collection order $\mv{I}$ to (P2) can be obtained via solving (P2-O) with optimized ${T}(I_{i},I_{i+1})$'s:
\begin{align}
\mbox{(P2-O)}\quad \underset{\mv{I}:\{I_i\}_{i=1}^N=\mathcal{N}}{\min}\quad  \sum_{i=0}^{N} {T}(I_{i},I_{i+1}).
\end{align}
\begin{figure}
\centering
\includegraphics[width=7cm]{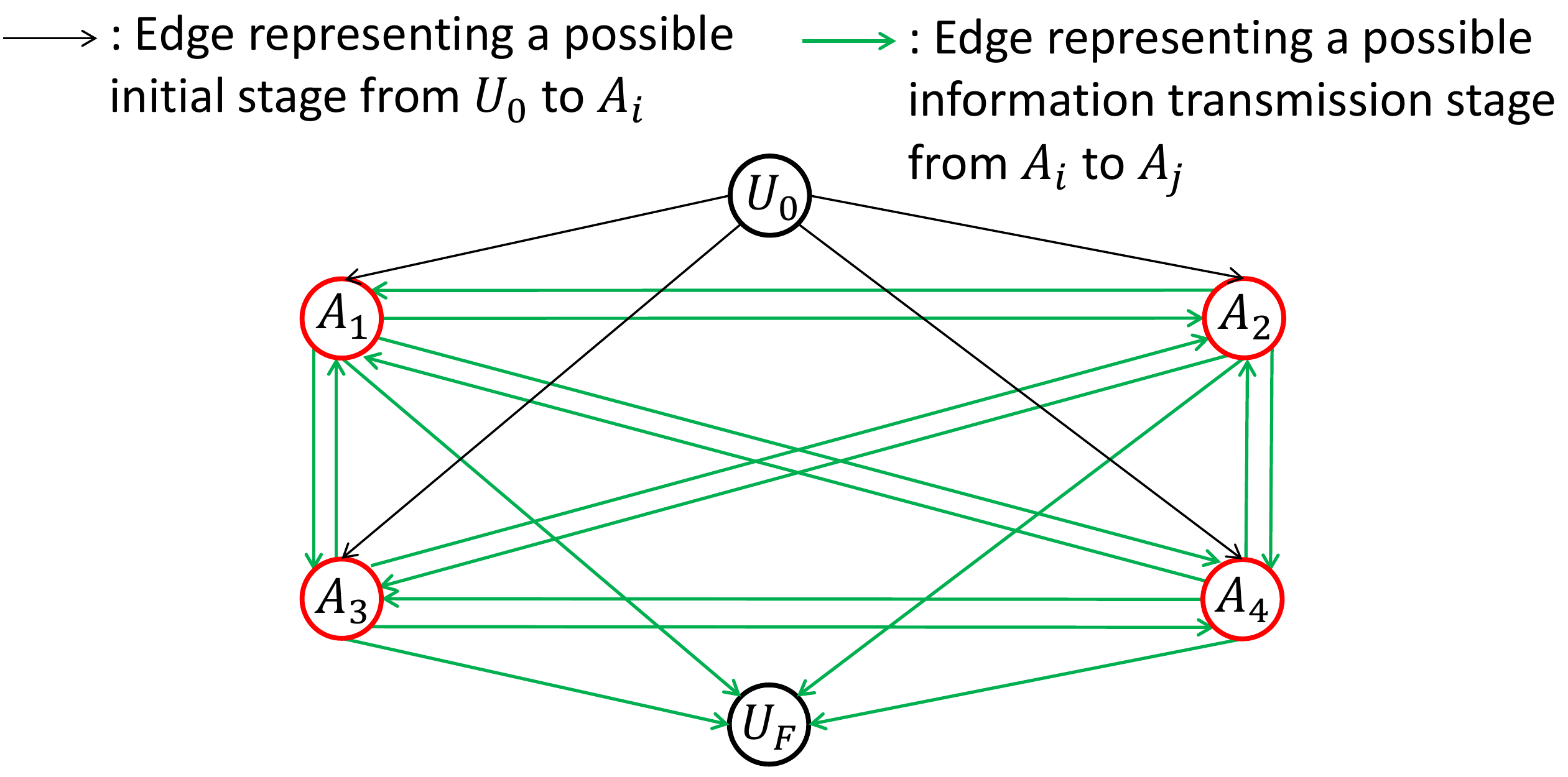}
\vspace{-3mm}
\caption{Illustration of graph $G=(V,E)$ with $N=4$.}	\label{graph}
\vspace{-3mm}
\end{figure}

Note that (P2-O) can be equivalently modeled as a TSP. Specifically, we construct a directed weighted graph denoted by $G=(V,E)$. The vertex set $V$ is given by 
\begin{equation}\label{vertex}
V=\{U_0,A_1,A_2,...,A_N,U_F\},
\end{equation}
where $U_0$ and $U_F$ represent the UAV's initial and final locations, respectively; $A_i$ represents the $i$-th information collection point. The edge set $E$ is given by 
\begin{align}\label{edge}
E=\{(U_0,A_i),(A_i,U_F),(A_i,A_j),i \neq j, i,j \in \mathcal{N}\},
\end{align}
where an edge exists between two points if they can be consecutively visited by the UAV during its mission, either in the initial stage or an information transmission stage, as illustrated in Fig. \ref{graph}. The weight of each edge is given by 
\begin{align}\label{weight}
W(U_0,A_i) = &{T}(0,i),\quad i \in \mathcal{N}\\[-1mm]
W(A_i,A_j) = &{T}(i,j),\quad i \neq j, i,j \in \mathcal{N}\\[-1mm]
W(A_i,U_F) = &{T}(i,N+1),\quad i \in \mathcal{N},
\end{align}
where $W(U_0,A_i)$ represents the time duration of the initial stage if $A_i$ is selected as the firstly visited information collection point; $W(A_i,A_j)$ and $W(A_i,U_F)$ represent the time durations required for the UAV to fly from $A_i$ to $A_j$ or $A_i$ to $U_F$, respectively, while finishing transmission of all the $D_i$-bit information for target $i$. Note that any \emph{path} in $G$ from $U_0$ to $U_F$ that visits all vertices denoted by $(U_0,A_{I_1},...,A_{I_N},U_F)$ corresponds to a \emph{feasible solution} of $\mv{I}=[I_1,...,I_N]^T$ to (P2-O), while the sum path weight is the objective value of (P2-O). 

Therefore, (P2-O) is equivalent to finding the \emph{shortest path} in graph $G$ from $U_0$ to $U_F$ that visits all vertices, which is the \emph{No-Return-Given-Origin-and-End} TSP \cite{TSP1}. Hence, the optimal solution to (P2) can be found by first obtaining the optimal $T(0,i)$, ${T}(i,j)$, and ${T}(i,N+1)$ for all $i,j\in \mathcal{N}$, $i\neq j$ via solving (P2-Sub), and then solving the TSP (P2-O) via exhaustively searching over all the feasible $\mv{I}$. 

However, such an optimal solution is difficult to be obtained in practice, because the exhaustive search method for (P2-O) incurs complexity $\mathcal{O}(N!)$, while (P2-Sub) is still a non-convex optimization problem due to the non-convex constraint in (\ref{P2subc_info}). To address these issues, we propose to find an approximate solution to the TSP (P2-O) with a low-complexity algorithm (e.g., the nearest-neighbour algorithm with complexity $\mathcal{O}(N^3)$), for which the details can be found in \cite{TSP1} and are omitted here due to limited space. Moreover, we will propose a high-quality suboptimal solution to (P2-Sub) by exploiting the problem structure, as shown in the following subsection.

\subsection{Proposed Solution to (P2-Sub)}
\begin{figure}[t]
\centering
\includegraphics[width=6.5cm]{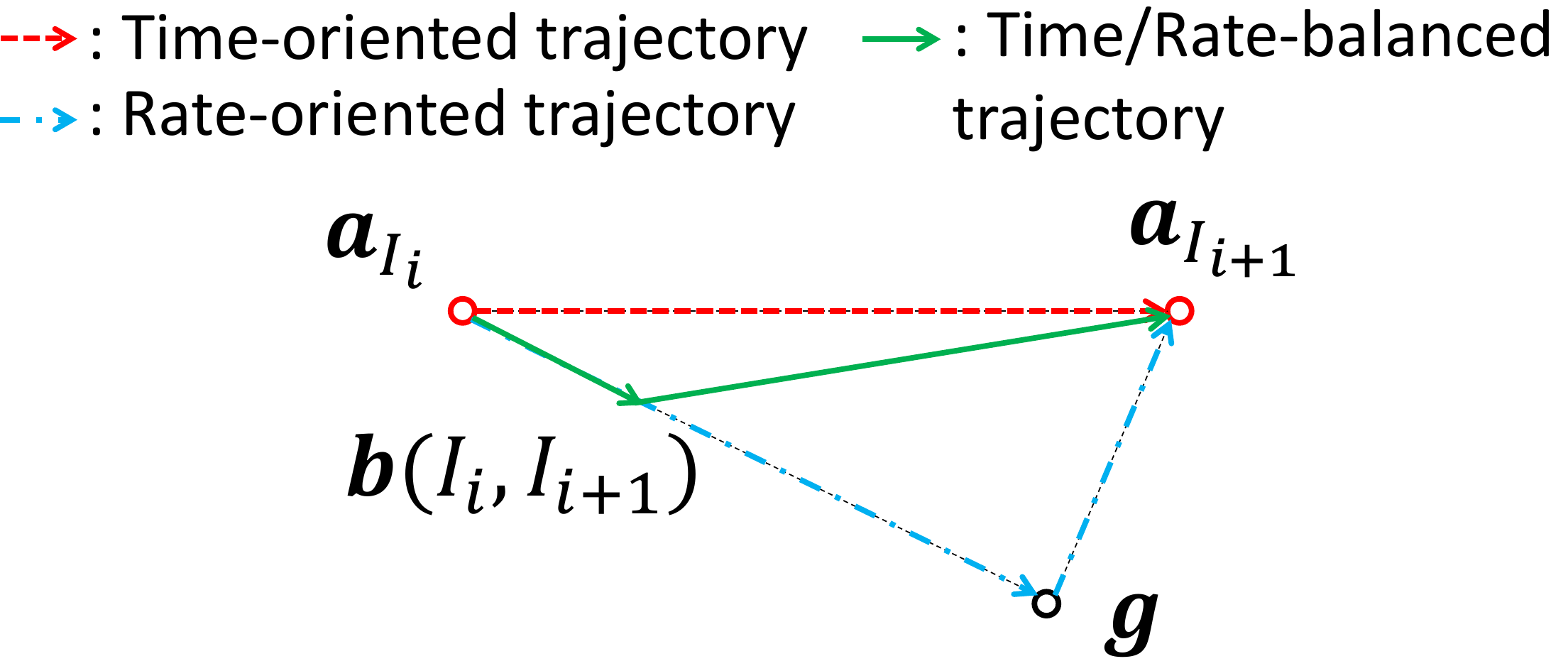}
\vspace{-3mm}
\caption{Illustration of the proposed trajectory structures.}\label{structure}
\vspace{-3mm}
\end{figure}

To start with, we derive a useful property of the optimal trajectory for (P2-Sub).
\begin{proposition}\label{proposition1}
The UAV's horizontal path corresponding to the optimal solution of (P2-Sub) must lie within a triangle with vertices $\mv{a}_{I_i}$, $\mv{g}$, and $\mv{a}_{I_{i+1}}$.
\end{proposition}
\begin{IEEEproof}
Please refer to Appendix A.
\end{IEEEproof}
Proposition \ref{proposition1} significantly reduces the feasible path range from the infinite space to a finite-sized triangle shown in Fig. \ref{structure}. Next, we propose three trajectory structures tailored for different information volume requirements within this triangle. 
\subsubsection{Time-Oriented Trajectory}
First, note that without the information transmission constraint in (\ref{P2subc_info}), the UAV's trajectory only needs to minimize the flying time, for which the optimal solution (termed as the \emph{time-oriented trajectory}) can be easily shown to be the straight-line flight from $\mv{a}_{I_i}$ to $\mv{a}_{I_{i+1}}$ with maximum speed $V_{\max}$, i.e.,
\begin{equation}\label{traj_time}
\bar{\mv{u}}_T(t)\!=\!\boldsymbol{a}_{I_i}\!+\!\frac{V_{{\rm max}}(\boldsymbol{a}_{I_{i+1}}\!-\!\boldsymbol{a}_{I_i})t}{\|\boldsymbol{a}_{I_{i+1}}\!-\!\boldsymbol{a}_{I_i}\|}, t\in[0,{T}_T(I_i,I_{i+1})],
\end{equation}
where ${T}_T(I_i,I_{i+1})=\frac{\|\boldsymbol{a}_{I_{i+1}}-\boldsymbol{a}_{I_i}\|}{V_{{\rm max}}}$ is the minimum flying time from $\mv{a}_{I_i}$ to $\mv{a}_{I_{i+1}}$. The total volume of information transmitted to the GBS with the time-oriented trajectory is $D_T(I_i,I_{i+1}) = \int_{0}^{T_T(I_i,I_{i+1})}R(\bar{\mv{u}}_T(t))dt$. Thus, it follows directly that if $D_{I_i}\leq D_T(I_i,I_{i+1})$, the optimal solution to (P2-Sub) is achieved by the time-oriented trajectory shown in (\ref{traj_time}), since ${T}_T(I_i,I_{i+1})$ is a lower bound of ${T}(I_i,I_{i+1})$.\footnote{An example of this case is the trajectory optimization in the initial stage, i.e., (P2-Sub) with $i=0$, where $D_{I_0}=0$ and (\ref{traj_time}) is always optimal.}

\subsubsection{Rate-Oriented Trajectory}
Then, we consider a \emph{rate-oriented trajectory} aiming to achieve high rate in the flight. Note from (\ref{rate}) that the UAV's transmission rate increases as the horizontal location of the UAV approaches that of the GBS, $\mv{g}$. Motivated by this, we let the UAV horizontally fly from $\mv{a}_{I_i}$ to $\mv{g}$ in straight line with speed $V_{\max}$, hover at $\mv{g}$ for time $\bar{T}\geq 0$, and then fly to $\mv{a}_{I_{i+1}}$ in straight line \hbox{with speed $V_{\max}$, i.e.,}
\begin{align}\label{maxpath}
&\bar{\mv{u}}_{R}(t,\bar{T}) = 
\\
&\begin{cases}
	\boldsymbol{a}_{I_i}\!+\!\frac{V_{{\rm max}}(\boldsymbol{g}-\boldsymbol{a}_{I_i})t}{\|\boldsymbol{g}-\boldsymbol{a}_{I_i}\|}, &t\in[0,T_g]\\
	\mv{g}, &t\in [T_g,T_g+\bar{T}]\\
	\boldsymbol{g}\!+\!\frac{V_{{\rm max}}(\boldsymbol{a}_{I_{i+1}}-\boldsymbol{g})(t-T_g-\bar{T})}{\|\boldsymbol{a}_{I_{i+1}}-\boldsymbol{g}\|}, &t\in[T_g+\bar{T},{T}_R(I_i,I_{i+1})],
\end{cases}\nonumber
\end{align}
with $T_g=\frac{\|\boldsymbol{a}_{I_i}-\boldsymbol{g}\|}{V_{{\rm max}}}$ and ${T}_R(I_i,I_{i+1},\bar{T})=T_g+\bar{T}+\frac{\|\boldsymbol{g}-\boldsymbol{a}_{I_{i+1}}\|}{V_{{\rm max}}}$. The total transmitted information volume is $D_R(I_i,I_{i+1},\bar{T})= \int_{0}^{T_R(I_i,I_{i+1},\bar{T})}R(\bar{\mv{u}}_R(t))dt$. Note that $D_R(I_i,I_{i+1},\bar{T})\geq D_R(I_i,I_{i+1},0), \forall \bar{T}$. Therefore, when $D_{I_i}\geq D_R(I_i,I_{i+1},0)$, a feasible solution to (P2-Sub) can always be found via (\ref{maxpath}), where the minimum required hovering time $\bar{T}$ is given by
\begin{align}
\bar{T}=\frac{D_{I_i}- D_R(I_i,I_{i+1},0)}{R(\mv{g})}=\frac{D_{I_i}- D_R(I_i,I_{i+1},0)}{B\log_2(1+\frac{P\beta_0}{H^2\sigma^2})}.
\end{align}

\subsubsection{Time/Rate-Balanced Trajectory}\label{bisection}
Note that the time-oriented trajectory and rate-oriented trajectory are suitable for the cases with low information volume $D_{I_i}\leq D_T(I_i,I_{i+1})$ and high information volume $D_{I_i}\geq D_R(I_i,I_{i+1},0)$, respectively. In the following, we consider the case with a moderate information volume $D_{I_i}\in(D_T(I_i,I_{i+1}),D_R(I_i,I_{i+1},0))$, and propose a novel trajectory structure for (P2-Sub) which balances the considerations on the flying time and transmission rate. To start with, we have the following proposition.
\begin{proposition}\label{maxvelocity}
If $D_{I_i}\in(D_T(I_i,I_{i+1}),D_R(I_i,I_{i+1},0))$, the optimal trajectory for (P2-Sub) has a constant speed $V_{\max}$.
\end{proposition}
\begin{IEEEproof}
Please refer to Appendix B.
\end{IEEEproof}

Based on the optimal trajectory properties revealed in Propositions \ref{proposition1} and \ref{maxvelocity}, we propose a \emph{time/rate-balanced trajectory} illustrated in Fig. \ref{structure}. Specifically, the UAV first (horizontally) flies from $\mv{a}_{I_i}$ to a location $\mv{b}(I_i,I_{i+1})$ on the line segment between $\mv{a}_{I_i}$ and $\mv{g}$, and then flies from $\mv{b}(I_i,I_{i+1})$ to $\mv{a}_{I_{i+1}}$, in straight lines with maximum speed, i.e.,
\begin{align}\label{balanced}
\bar{\mv{u}}_{B}(t) = 
\begin{cases}
	\boldsymbol{a}_{I_i}\!+\!\frac{V_{{\rm max}}(\mv{b}(I_i,I_{i+1})-\boldsymbol{a}_{I_i})t}{\|\mv{b}(I_i,I_{i+1})-\boldsymbol{a}_{I_i}\|},\quad t\in[0,T_b]\\
	\mv{b}(I_i,I_{i+1})\!+\!\frac{V_{{\rm max}}(\boldsymbol{a}_{I_{i+1}}-\mv{b}(I_i,I_{i+1}))(t-T_b)}{\|\boldsymbol{a}_{I_{i+1}}-\mv{b}(I_i,I_{i+1})\|},\\
	\qquad\qquad\qquad\qquad t\in[T_b,{T}_B(I_i,I_{i+1})],
\end{cases}\!\!\!\!
\end{align}
with $T_b\!\!=\!\!\frac{\|\mv{b}(I_i,I_{i+1})\!-\!\boldsymbol{a}_{I_i}\|}{V_{\max}}$, ${T}_B(I_i,I_{i+1})\!\!=\!\!T_b\!+\!\frac{\|\boldsymbol{a}_{I_{i+1}}\!-\!\mv{b}(I_i,I_{i+1})\|}{V_{\max}}$.

It can be shown that as $\mv{b}(I_i,\!I_{i+1})$ moves from $\mv{a}_{I_i}$ to $\mv{g}$, the flying time increases, while the transmitted information volume also increases since the transmission rate increases as the UAV-GBS distance decreases. Particularly, this trajectory reduces to the time-oriented trajectory with $D_T(I_i,\!I_{i+1})$ when $\mv{b}(I_i,\!I_{i+1})\!=\!\mv{a}_{I_i}$, and the rate-oriented trajectory with $\bar{T}\!=\!0$ and $D_R(I_i,\!I_{i+1},\!0))$ when $\mv{b}(I_i,\!I_{i+1})\!=\!\mv{g}$. Thus, the proposed trajectory can achieve a flexible \emph{trade-off} between flying time and information volume by tuning $\mv{b}(I_i,\!I_{i+1})$. For given $D_{I_i}$, the optimal $\mv{b}(I_i,\!I_{i+1})$ can be efficiently obtained via \emph{one-dimensional bi-section search} over the line segment between $\mv{a}_{I_i}$ and $\mv{g}$, until the transmitted information volume $D_B(I_i,I_{i+1}) = \int_{0}^{T_B(I_i,I_{i+1})}R(\bar{\mv{u}}_B(t))dt$ equals to $D_{I_i}$.

Hence, the proposed trajectory for (P2-Sub) is given by
\vspace{0.5mm}
\begin{align}\label{traj}
\!\!\!	\bar{\mv{u}}(t) = 
\begin{cases}
	\bar{\mv{u}}_T(t), D_{I_i}\in [0,D_T(I_i,I_{i+1})]\\[-1mm]
	\bar{\mv{u}}_B(t),D_{I_i}\in(D_T(I_i,I_{i+1}),D_R(I_i,I_{i+1},0))\\[-1mm]
	\bar{\mv{u}}_R(t,\bar{T}),D_{I_i}\in[D_R(I_i,I_{i+1},0),\infty).
\end{cases}\!\!\!\!\!
\end{align}
\vspace{1mm}
The corresponding objective value of (P2-Sub) is given by
\begin{align}\label{T}
&T(I_i,I_{i+1}) = \\
&	\begin{cases}
	{T}_T(I_i,I_{i+1}), &D_{I_i}\in [0,D_T(I_i,I_{i+1})]\\[-1mm]
	{T}_B(I_i,I_{i+1}),&D_{I_i}\in(D_T(I_i,I_{i+1}),D_R(I_i,I_{i+1},0))\\[-1mm]
	{T}_R(I_i,I_{i+1},\bar{T}),&D_{I_i}\in[D_R(I_i,I_{i+1},0),\infty).
\end{cases}\!\!\!\!\!\nonumber
\end{align}
\subsection{Summary of the Overall Algorithm for (P2)}
Finally, we summarize the overall algorithm for (P2). First, we obtain $T(0,i)$, ${T}(i,j)$, and ${T}(i,N+1)$ based on (\ref{T}) for all $i,j\in \mathcal{N},i\neq j$, as well as the corresponding sub-trajectory solutions $\{\bar{\mv{u}}(t)\}$'s based on (\ref{traj}). Then, we construct the graph $G$ based on (\ref{vertex})-(\ref{weight}), and apply the TSP algorithm in \cite{TSP1} with complexity $\mathcal{O}(N^3)$ to find an approximate solution of $\mv{I}$ to (P2-O). Finally, the proposed solution to (P2) is obtained via (\ref{P2sol_1}) and (\ref{P2sol_2}), which is guaranteed to be a feasible solution for both (P2) and (P1). The worst-case complexity of our proposed algorithm can be shown to be $\mathcal{O}(N^3+N^2\log(\underset{I_i}{\max}\|\mv{a}_{I_i}-\mv{g}\|/\epsilon))$, where $\epsilon$ characterizes the precision requirement of the bi-section search in Section \ref{bisection}.

\section{Numerical Results}
\begin{figure}[t]
\centering
\subfigure{
	{\scalebox{0.31}{\includegraphics*{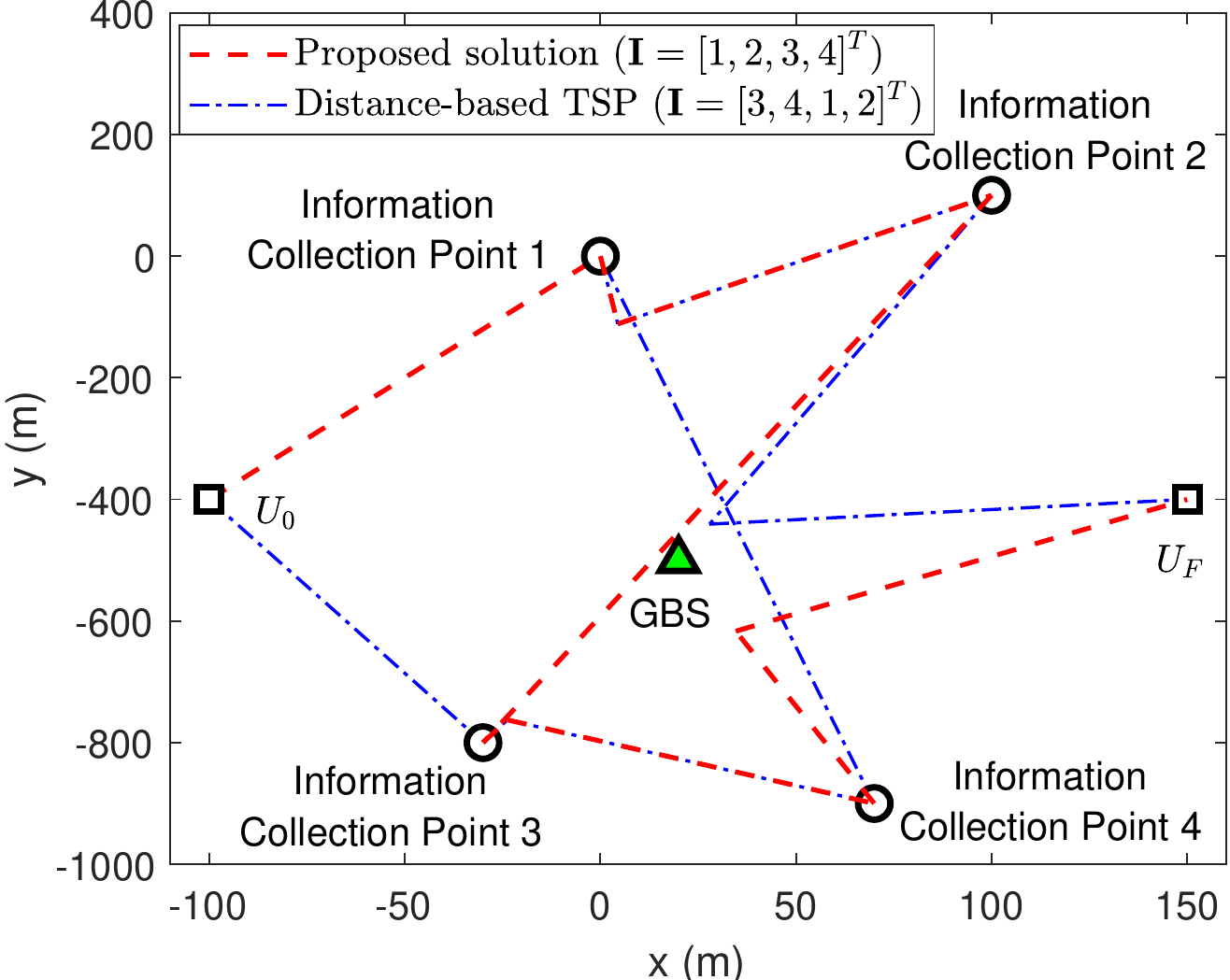}}}}
\subfigure{
	{\scalebox{0.31}{\includegraphics*{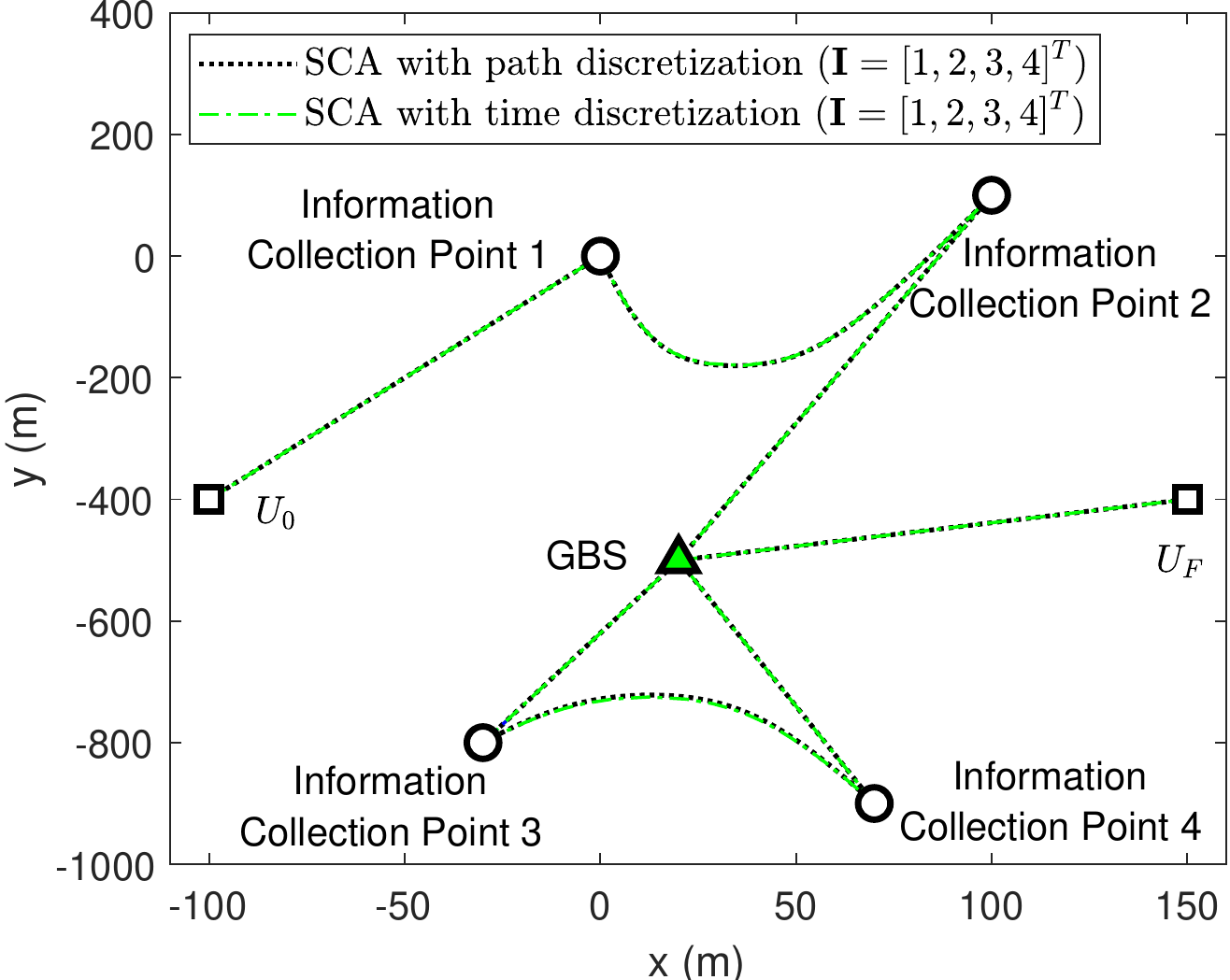}}}}
\vspace{-3mm}
\caption{Trajectory comparison for different schemes under $\alpha=1$.}\label{trajectory}
\vspace{-3mm}\end{figure}

In this section, we provide numerical results to evaluate the performance of our proposed trajectory design. We set $B\!=\!1$ MHz, $P\!=\!30$ dBm, $\beta_0\!=\!-60$ dB, $\sigma^2\!=\!-110$ dBm, and $V_{\max}\!=\!10$ m/s. The UAV's altitude is fixed at $H_U\!=\!120$ m, and the GBS's height is set as $H_G\!=\!20$ m, thus $H\!=\!H_U\!-\!H_G=100$ m. We consider $N\!=\!4$ targets, whose horizontal locations are $\mv{a}_1\!=\![0,0]^T$, $\mv{a}_2\!=\![100,100]^T$, $\mv{a}_3\!=\![-30,-800]^T$, $\mv{a}_4\!=\![70,-900]^T$, respectively, as illustrated in Fig. \ref{trajectory}. The horizontal locations of the GBS, $U_0$, and $U_F$ are $\mv{g}\!=\![20,-500]^T$, $\mv{u}_0\!=\![-100,-400]^T$, and $\mv{u}_F\!=\![150,-400]^T$, respectively. Denote $\{D_i\!=\!\alpha \bar{D}_i\}_{i=1}^4$ as the required information volume for the $4$ targets, where $\alpha$ is a scaling factor; $\bar{D}_1\!=\!3\times 10^8$, $\bar{D}_2\!=\!7\times 10^8$, $\bar{D}_3\!=\!2\times 10^8$, and $\bar{D}_4\!=\!6\times 10^8$ are a set of reference volume. For comparison, we consider the following benchmark schemes:
\begin{itemize}[leftmargin=*]
\item {\bf{Distance-based TSP}}: In this scheme, we design the information collection order $\mv{I}$ to minimize the total flying distance from $U_0$ to $U_F$ while visiting all information collection points via the TSP algorithm \cite{TSP1}. Based on the obtained $\mv{I}$, the trajectory is designed based on our proposed path structure according to (\ref{traj}), (\ref{P2sol_1}), and (\ref{P2sol_2}).
\item {\bf{SCA with path/time discretization}}: Note that (P2-Sub) can also be handled via the SCA method, by first approximating the continuous trajectory with discrete line segments, and then optimizing the end points of them via successively solving a convex approximation of (P2-Sub). In these schemes, we adopt the SCA method with path or time discretization \cite{Access} to find a suboptimal solution of (P2-Sub), based on which $\mv{I}$ is designed via the TSP algorithm over graph $G$.
\end{itemize}
\begin{figure}[t]
\centering
\includegraphics[width=6cm]{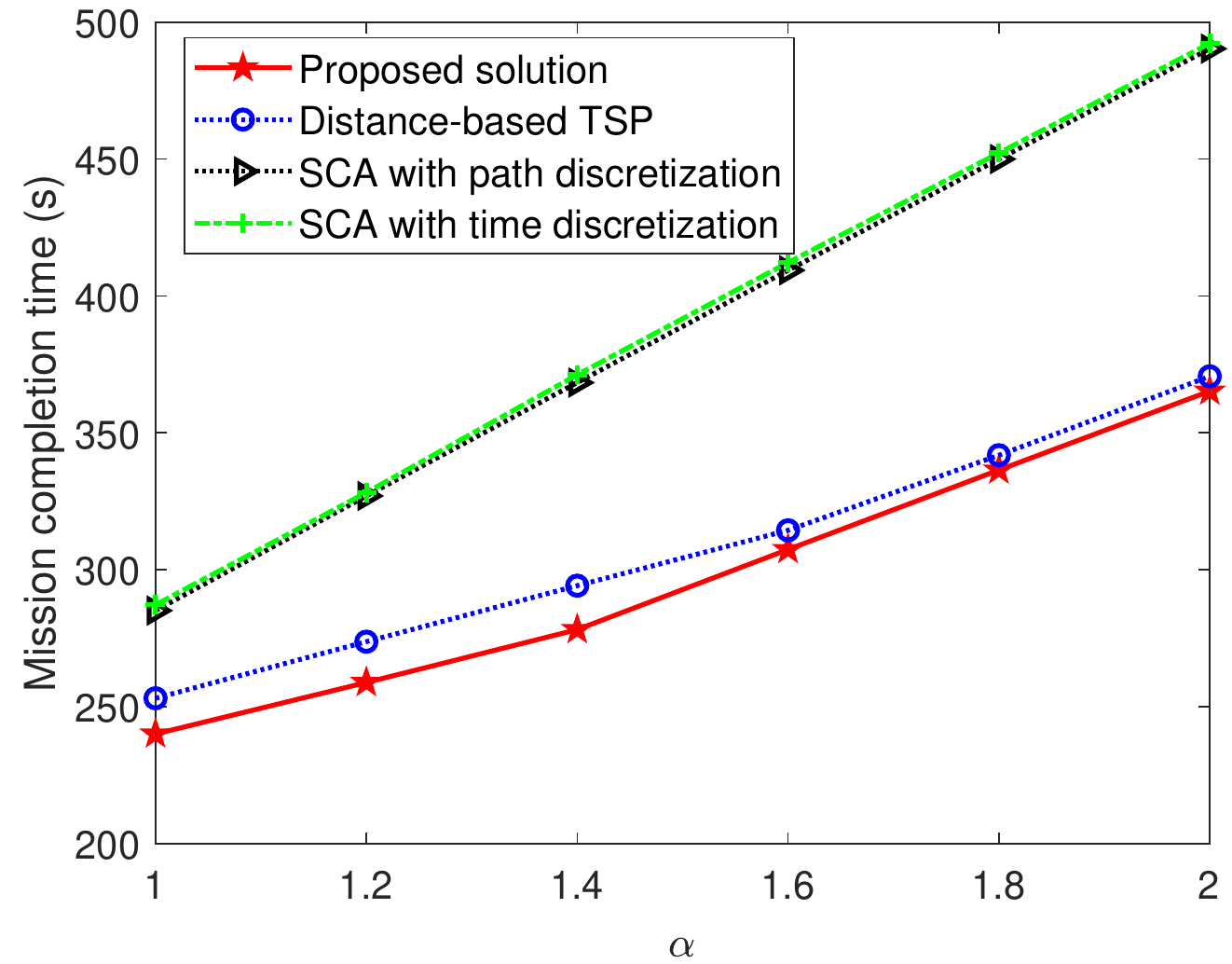}
\vspace{-4mm}
\caption{Mission completion time comparison for different schemes.}\label{completion}
\vspace{-3mm}\end{figure}

In Fig. \ref{completion}, we show the mission completion time $T$ for different schemes versus $\alpha$, where we set $T_i^{\cc}\!=\!0,\forall i$ for simplicity. It is observed that the proposed solution requires the minimum mission completion time among all schemes for all values of $\alpha$, since our judicious design of the information collection order and UAV's sub-trajectories by exploiting the unique problem structures is able to well balance between the required time and transmission rate. In Fig. \ref{trajectory}, we show the trajectories of different schemes for $\alpha=1$. It is observed that the information collection order designed based on graph $G$ in (\ref{vertex})-(\ref{weight}) for both our proposed solution and the SCA-based methods is $\mv{I}\!=\![1,2,3,4]^T$, while that for distance-based TSP is $\mv{I}\!=\![3,4,1,2]^T$. Note that the latter is unfavorable since the information of target $2$ with largest required  volume is lastly collected, thus requiring long time duration for the last information transmission stage. On the other hand, the SCA-based trajectories generally have curved paths of longer lengths compared to those in the proposed solution and distance-based TSP. Moreover, it can be observed from Fig. \ref{completion} that the performance gap between the SCA-based methods and our proposed solution increases as the required information volume increases, due to the inaccuracies in the problem approximation and trajectory discretization. This further demonstrates the superiority of our proposed trajectory structure tailored for the considered problem.

\vspace{-1mm}
\section{Conclusion}
\vspace{-1mm}
In this paper, we studied a cellular-connected UAV which needs to hover at multiple fixed locations for information collection, and transmit the collected information to the GBS during its flight. We formulated the joint optimization problem of the UAV's trajectory and the information collection order to minimize the mission completion time, which is an NP-hard problem. By devising structured communication protocols, we reformulated the problem into a more tractable form, and established a graph theory based method for finding a high-quality suboptimal solution. Numerical results showed that our proposed solution outperforms various benchmark schemes.

\begin{appendices}
\section{Proof of Proposition \ref{proposition1}}\label{A}
\begin{figure}[t]
	\centering
	\includegraphics[width=9cm]{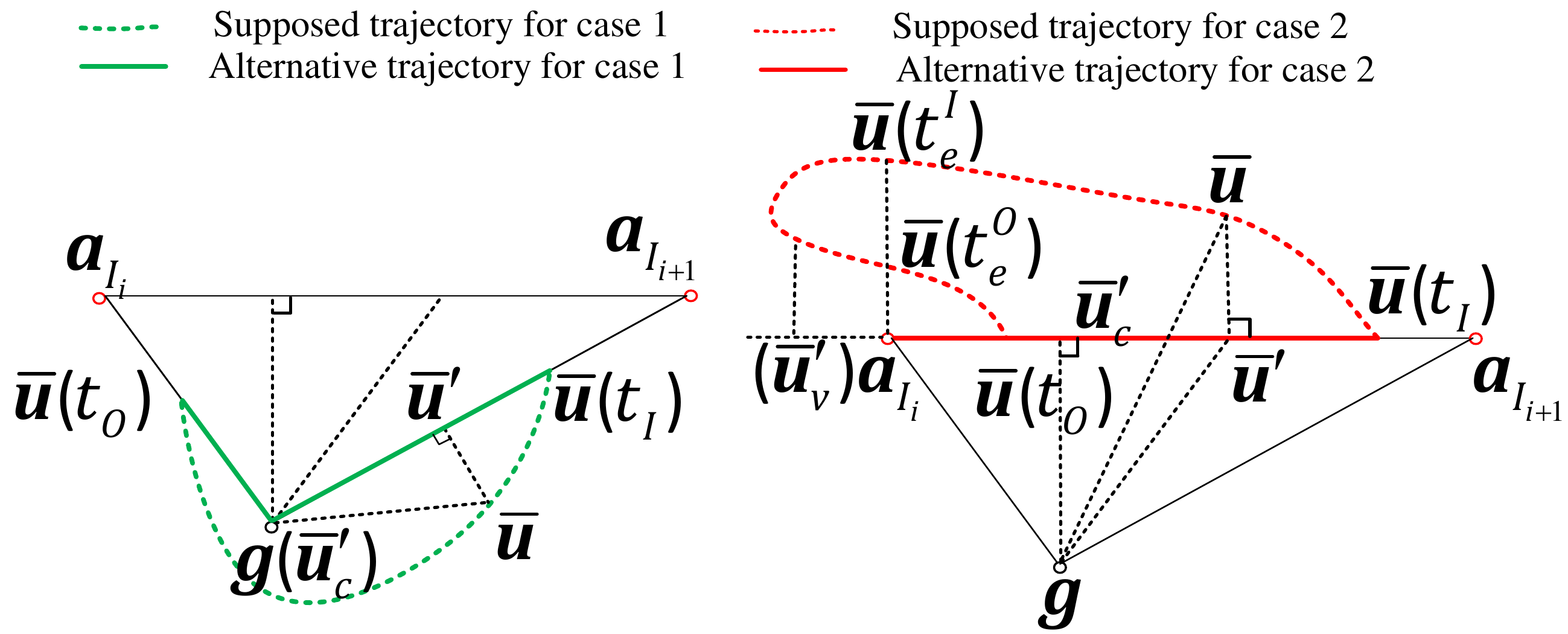}
	\vspace{-5mm}
	\caption{Illustration for the proof of Proposition~\ref{proposition1}.}\label{proof1}
	\vspace{-3mm}\end{figure}

Consider a UAV trajectory solution to problem (P2-Sub) denoted by $\{\bar{\boldsymbol{u}}(t),t \in [0,T(I_i,I_{i+1})]\}$ with trajectory component $\{\bar{\boldsymbol{u}}(t),t \in (t_I,t_O), 0< t_I < t_O < T(I_i,I_{i+1})\}$ outside the triangle. Note that the locations at time instants $t_I$ and $t_O$, i.e., $\bar{\boldsymbol{u}}(t_I)$ and $\bar{\boldsymbol{u}}(t_O)$, lie on the triangle's edges, as illustrated in Fig.~\ref{proof1}. Then, we prove Proposition~\ref{proposition1} by showing that for any trajectory component $\{\bar{\boldsymbol{u}}(t),t \in (t_I,t_O)\}$ outside the triangle, we can always construct
an alternative trajectory, denoted by $\{\bar{\boldsymbol{u}}'(t),t \in (t_I,t_O)\}$, that lies within the triangle and has a larger information transmission volume. 

First, we observe that for any location $\bar{\boldsymbol{u}} \in \mathbb{R}^{2 \times 1}$ on the trajectory component, its projection on the triangle's edges (or on the extended lines of the triangle's edges), denoted by $\bar{\boldsymbol{u}}' \in \mathbb{R}^{2 \times 1}$, always has a smaller distance to the GBS, and consequently a higher transmission rate. According to this, we construct the alternative trajectory by considering the following two cases. 

\textit{Case 1}: All the projected locations of the trajectory lie on the edges of the triangle. Denote the projected location which is closest to the GBS as $\bar{\boldsymbol{u}}_c' \in \mathbb{R}^{2 \times 1}$. In this case, the alternative path consists of all these projections. Specifically, the UAV is set to fly from $\bar{\boldsymbol{u}}(t_I)$ to $\bar{\boldsymbol{u}}_c'$ in straight-line path with maximum speed $V_{{\rm max}}$, then hover over $\bar{\boldsymbol{u}}_c'$ for a time period $T_h$, and finally fly from $\bar{\boldsymbol{u}}_c'$ to $\bar{\boldsymbol{u}}(t_O)$ in straight-line path with maximum speed $V_{{\rm max}}$. The alternative trajectory is expressed as 
\begin{equation}
	\bar{\boldsymbol{u}}'(t) = \begin{cases}
		\bar{\boldsymbol{u}}(t_I)+\frac{V_{{\rm max}}(\bar{\boldsymbol{u}}_c'
			-\bar{\boldsymbol{u}}(t_I))t}{\|\bar{\boldsymbol{u}}_c'
			-\bar{\boldsymbol{u}}(t_I)\|}, t \in [t_I,t_I +T_{u,c}]\\
		\bar{\boldsymbol{u}}_c',\qquad\qquad\qquad   t \in (t_I+T_{u,c},t_I +T_{u,c}+T_h]\\       
		\bar{\boldsymbol{u}}_c'+\frac{V_{{\rm max}}(\bar{\boldsymbol{u}}(t_O)-\bar{\boldsymbol{u}}_c')(t-(t_I +T_{u,c}+T_h))}{\|
			\bar{\boldsymbol{u}}(t_O)-\bar{\boldsymbol{u}}_c'\|},\\
		\qquad\qquad\qquad\qquad\qquad t \in (t_I +T_{u,c}+T_h,t_O],	 		
	\end{cases}
\end{equation}    
where $T_{u,c} = \frac{\|\bar{\boldsymbol{u}}_c' - \bar{\boldsymbol{u}}(t_I)\|}{V_{{\rm max}}}$ is the time for the UAV to fly from $\bar{\boldsymbol{u}}(t_I)$ to $\bar{\boldsymbol{u}}_c'$, $T_h = t_O - t_I - T_{u,c} - \frac{{\|\bar{\boldsymbol{u}}(t_O)-\bar{\boldsymbol{u}}_c'\|}}{V_{{\rm max}}}$ is the time for UAV's hovering over $\bar{\boldsymbol{u}}_c'$. 

\textit{Case 2}: Some projected locations of the trajectory lie on the extended lines of the triangle's edges. Assume that the projected locations from time instant $t_e^I$ to $t_e^O$, $t_I\leq t_e^I < t_e^O \leq t_O$, lie on the extended lines of the triangle's edges, as shown in the right-hand side of Fig.~\ref{proof1}. During this time period, the UAV is not set to fly over the projections since they are outside the triangle. Instead, it is set to hover over the nearest vertex of the triangle, denoted by $\bar{\boldsymbol{u}}_v'$ which has shorter distance to the GBS than all the projected locations on the extended lines of the triangle's edges, thus yielding higher information transmission rate. The rest components of the alternative trajectory are constructed similarly as in \textit{Case 1}. In this case, the alternative trajectory is expressed as
\begin{equation}
	\bar{\boldsymbol{u}}'(t) = \begin{cases}
		\bar{\boldsymbol{u}}(t_I)+\frac{V_{{\max}}(\bar{\boldsymbol{u}}_c'
			-\bar{\boldsymbol{u}}(t_I))t}{\|\bar{\boldsymbol{u}}_c'
			-\bar{\boldsymbol{u}}(t_I)\|}, \qquad t \in [t_I,t_I +T_{u,c}]\\
		\bar{\boldsymbol{u}}_c', \qquad\qquad\qquad\  t \in (t_I+T_{u,c},t_I +T_{u,c}+T_h]\\    
		\bar{\boldsymbol{u}}_c'+\frac{V_{{\rm max}}(\bar{\boldsymbol{u}}_v'-\bar{\boldsymbol{u}}_c')(t-(t_I +T_{u,c}+T_h))}{\|\bar{\boldsymbol{u}}_v'-\bar{\boldsymbol{u}}_c'
			\|}, \\
		\qquad\qquad\qquad\qquad t \in (t_I +T_{u,c}+T_h,t_{e}^I]\\  
		\bar{\boldsymbol{u}}_v', \ \qquad\qquad\qquad  t \in (t_{e}^I,t_{e}^O]\\  
		\bar{\boldsymbol{u}}_v'+\frac{V_{{\rm max}}(\bar{\boldsymbol{u}}(t_O)-\bar{\boldsymbol{u}}_v')(t-t_{e}^O)}{\|
			\bar{\boldsymbol{u}}(t_O)-\bar{\boldsymbol{u}}_v'\|}, \quad t \in (t_{e}^O,t_O],	 		
	\end{cases}
\end{equation}
where the time for hovering over $\bar{\boldsymbol{u}}_c'$ is derived as $T_h = t_O - t_I -T_{u,c} - \frac{{\|\bar{\boldsymbol{u}}_c'-\bar{\boldsymbol{u}}_v'\|}}{V_{{\rm max}}} - (t_e^O - t_e^I) - \frac{{\|\bar{\boldsymbol{u}}(t_O)-\bar{\boldsymbol{u}}_v'\|}}{V_{{\rm max}}}$, with $\frac{{\|\bar{\boldsymbol{u}}(t_O)-\bar{\boldsymbol{u}}_v'\|}}{V_{{\rm max}}}$ being the time for flying from $\bar{\boldsymbol{u}}(t_O)$ to $\bar{\boldsymbol{u}}_v'$. 

Based on the above alternative trajectory construction, for any location at any time instant on the original trajectory, we can always find a corresponding location at one time instant on the alternative trajectory that has a lower distance to the GBS, and consequently a larger information transmission rate. Hence, the alternative trajectory always yields a larger information transmission volume. This thus completes the proof of Proposition \ref{proposition1}.

\section{Proof of Proposition \ref{maxvelocity}}\label{B}
We prove Proposition 2 by showing that with $D_{I_i}\in(D_T(I_i,I_{i+1}),D_R(I_i,I_{i+1},0))$, for any feasible solution to (P2-Sub) denoted by $\{\bar{\boldsymbol{u}}(t),  t \in [0,T(I_i,I_{i+1})]\}$ where the UAV's speed is less than $V_{\max}$ over certain time periods, we can always construct a new feasible solution $\{\bar{\boldsymbol{u}}'(t), t \in [0,T(I_i,I_{i+1})]\}$ with $\|\dot{\boldsymbol{u}}'(t)\| = V_{{\rm max}},\forall t$ and larger information transmission volume throughout the flight.

\begin{figure}
	\centering
	\includegraphics[width=6cm]{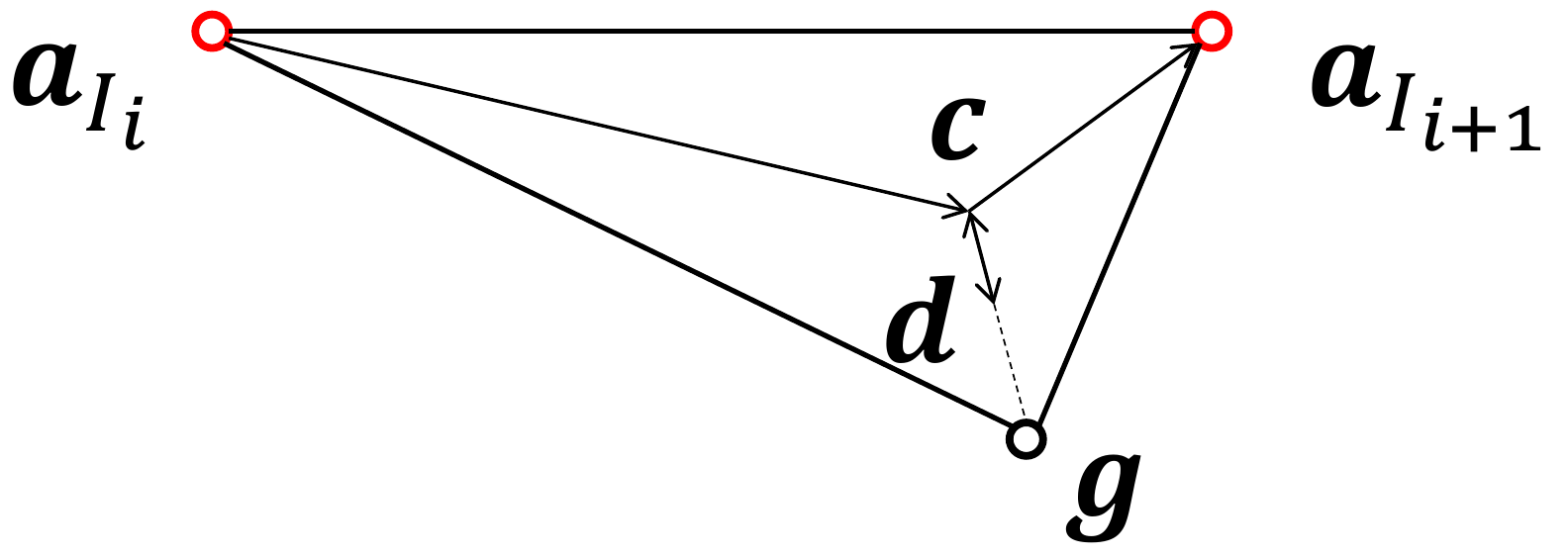}
	\caption{Illustration for the proof of Proposition 2.}
	\label{speed}
\end{figure}

Specifically, we let $\mv{c}\in \mathbb{R}^{2\times 1}$ denote the location that is closest to $\mv{g}$ among all the UAV's possible locations in $\bar{\boldsymbol{u}}(t)$. The path of the new trajectory $\{\bar{\boldsymbol{u}}'(t), t \in [0,T(I_i,I_{i+1})]\}$ consists of three components: the path from $\mv{a}_{I_i}$ to $\mv{c}$ in the original path, the straight-line path from $\mv{c}$ back-and-forth to another location $\mv{d}\in \mathbb{R}^{2\times 1}$ on the line segment between $\mv{c}$ and $\mv{g}$ and finally getting back to $\mv{c}$, and the path from $\mv{c}$ to $\mv{a}_{I_{i+1}}$ in the original path, as illustrated in Fig. \ref{speed}.\footnote{Note that since $D_{I_i}\in(D_T(I_i,I_{i+1}),D_R(I_i,I_{i+1},0))$, it can be shown that the optimal trajectory does not need to traverse $\mv{g}$ as in the rate-oriented trajectory in (\ref{maxpath}), since otherwise it will incur longer flying time. Hence, we have $\mv{c}\neq \mv{g}$, and consequently $\mv{d}$ exists.} The UAV is assumed to fly with maximum speed $V_{\max}$ throughout its entire flight. Note that in the newly constructed trajectory, the total required time to fly from $\mv{a}_{I_i}$ to $\mv{c}$ and from $\mv{c}$ to $\mv{a}_{I_{i+1}}$ is $T'=\frac{\|\mv{a}_{I_i}-\mv{c}\|+\|\mv{a}_{I_{i+1}}-\mv{c}\|}{V_{\max}}$. The location of $\mv{d}$ is then determined such that flying between $\mv{c}$ and $\mv{d}$ back-and-forth in straight-line with speed $V_{\max}$ and finally reaching $\mv{c}$ takes time $T(I_i,I_{i+1})-T'$. Note that compared to the locations on the original path, every location on the newly constructed path between $\mv{c}$ and $\mv{d}$ has a smaller distance to $\mv{g}$, as illustrated in Fig. \ref{speed}, thus yielding a larger transmission rate. Therefore, it can be shown that the newly constructed trajectory achieves a higher information transmission volume. Consequently, the optimal trajectory to (P2-Sub) when $D_{I_i}\in(D_T(I_i,I_{i+1}),D_R(I_i,I_{i+1},0))$ should have a constant maximum speed of $V_{\max}$. This thus completes the proof of Proposition 2.

\end{appendices}

\end{document}